\documentstyle[aps,epsf]{revtex}

\begin{document}

\title{Three-parameter model of a sand pile}

\author{Alexander I. Olemskoi}
\address{Department of Physical Electronics,
Sumy State University\\ 2, Rimskii-Korsakov St., 244007 Sumy,
UKRAINE\\
E--mail: Alexander@olem.sumy.ua}

\maketitle

\begin{abstract}
The theory of a flux steady-state (avalanche) formation
is presented for the simplest model of a real sand pile
within the framework of Lorenz approach.
The stationary values of sand velocity and
sand pile slope are derived as functions
of controlling parameter (externally driven sandpile slope).
The additive noises of above values
are taken into account to build the phase diagram,
where the noise intensities determine a
domain of the avalanche appearance.
This domain shows to be crucial to the noise intensity
of vertical component of sand velocity.

\end{abstract}



\section{Introduction}

In recent years considerable study has been given to the theory
of self--organized criticality (SOC) [1]
that explains spontaneous (avalanche-type) dynamics,
unlike the typical phase transitions that occur
only when a controlling parameter is tuned to a critical value.
A main feature of the systems displaying SOC is that they are distributed
over avalanche sizes, so that SOC models are
mostly studied by making use of scaling-type arguments supplemented
with extensive computer simulations (see [2]).
On the contrary, in this paper we put forward
the theory of a single avalanche formation.

The SOC behavior appears in a variety of systems
such as a real sand pile
(ensemble of grains of sand moving
along increasingly tilted surface),
intermittency in biological evolution [3],
earthquakes and propagation of forest--fires, depinning transitions
in random medium and so on [4].
Among the above models the simplest and most widely studied,
analytically [5], [6]
and numerically [1], [7], [8],
are the sandpile models.
The field theory [9], based on nonlinear diffusion equation, has
failed to account for the basic feature of self--organized
systems -- their avalanche dynamics. The reason is that there is no
feedback between open subsystem and thermostat within the framework
of the one--parameter approach employed in [9]. Recently the
two--parameter theories were set forth in [10,11], where the
thermostat degree of freedom was either controlling parameter [10],
or conjugate field [11]. In the mean field approximation, the
approach of [11] allows to obtain critical exponents governing
scaling behavior of self--organized system.
Our approach is to take into consideration the complete set of
degrees of freedom. Owing to this, not only the complimentary results
of [10,11] are reproduced, but it is possible to obtain the
self--consistent analytical description of the single avalanche formation.

This paper is organized as follows. In Sec. 2 the self-consistent theory
of the formation of a flux steady-state is presented. It enables us to treat
the problem on the basis of the unified analytical approach.
Sec. 3 deals with accounting additive noises of the sand velocity components
and sandpile slope. It is shown that the noise intensities increase
the possibility of emergence of the SOC regime.
Sec.4 contains the obtained results.

\section{Noiseless case}

Within framework of the simplest model of a sandpile, a dependence $y=y(t,x)$
defines its surface at given instant of time $t$. Locally the flow of
sand can be described in terms of three quantities: the horizontal
and vertical components of the sand velocity,
$\dot{x}=\partial x/ \partial t $,
$\dot{y}=\partial y/ \partial t $, and the surface slope
$y^{\prime}=\partial y/ \partial x $. The key point of our approach
is that the above degrees of freedom are assumed to be of dissipative
type, so that, when they are not coupled, their relaxation to the
steady state is governed by the Debye-type equations:
\begin{mathletters}
\label{1}
\begin{equation}
\frac{\text{d}\dot{x}}{\text{d}t}=-\frac{\dot{x}}{\tau_{x}},
\label{1a}
\end{equation}
\begin{equation}
\frac{\text{d}\dot{y}}{\text{d}t}=-\frac{\dot{y}}{\tau_{y}^{(0)}},
\label{1b}
\end{equation}
\begin{equation}
\frac{\text{d}y^{\prime}}{\text{d}t}=
\frac{y^{\prime}_{0}-y^{\prime}}{\tau_{S}},
\label{1c}
\end{equation}
\end{mathletters}
where $\tau_{x},\,\tau_{y}^{(0)}$ and $\tau_{S}$ are the relaxation
times of the velocity components and the slope, respectively.
Eqs.
(\ref{1a}) -- (\ref{1c})
imply the sand is at rest in the stationary state,
$\dot{x}=\dot{y}=0$ and the equilibrium slope
$y^{\prime}=y^{\prime}_{0}\neq 0$ plays the role of a controlling
parameter.

Since the motion of sand grain along different directions is not
independent, Eq.(\ref{1a}) should be changed by adding the term
$f=\dot{y}/ \gamma$ due to liquid friction force along the
$y$--axis ($\gamma$ is the kinetic coefficient). Then, we have

\begin{equation}
\tau_{x} \ddot{x}= -\dot{x}+a^{-1}\dot{y}
\label{2}
\end{equation}
where $a\equiv \gamma/ \tau_{x}$.
Note that, owing to the diffusion equation
$\dot{y}=Dy^{\prime\prime}$ ($D$ is the diffusion coefficient), the
friction force appears to be driven by the curvature of the
sandpile surface:

\begin{equation}
f=(D/ \gamma)y^{\prime\prime}.
\label{3}
\end{equation}
On the other hand, when $\ddot{x}=0$ (stationary state), solution
of Eq.(\ref{2}) defines the tangent line $y=ax+const$, so that the
friction force  $f=\tau^{-1}_{x}\dot{x}$
is proportional to the horizontal component
of the sand velocity.
Taking into consideration
the relation (\ref{3}) and using the chain rule
$\text{d}y^{\prime}/\text{d}t=
\dot{y}^{\prime}+y^{\prime\prime}\dot{x}$,
from Eq.(\ref{1c})
one obtains the equation of motion for the slope:

\begin{equation}
\tau_{S}\dot{y}^{\prime}=(y^{\prime}_{0}-y^{\prime})-
\left(\tau_{S}/D \right)\dot{y}\dot{x}.
\label{4}
\end{equation}
Following the same line, the equation for the vertical
component of the velocity can be deduced

\begin{equation}
\tau_{y}\ddot{y}=-\dot{y}+
\frac{\tau_{y}}{\tau_{x}}y^{\prime}\dot{x},\qquad
\frac{1}{\tau_{y}}\equiv\frac{1}{\tau^{(0)}_{y}}
\left(
1+\frac{y^{\prime}_{0}}{a}\frac{\tau^{(0)}_{y}}{\tau_{x}}
\right).
\label{5}
\end{equation}
Note the higher order terms are disregarded in Eq.(\ref{5}) and the
renormalized relaxation time $\tau_{y}$ depending on the
stationary slope $y^{\prime}_{0}$ is introduced.

Eqs.(\ref{2}), (\ref{4}), (\ref{5}) constitute the basis for
self--consistent description of the sand flow on the surface with the
slope $y^{\prime}$ driven by the control parameter $y^{\prime}_{0}$.
The distinguishing feature of these equations is that nonlinear terms
that enter Eqs.(\ref{4}), (\ref{5}) are of opposite signs, while
Eq.(\ref{2}) is linear. Physically, the latter means that on the
early stage the avalanche begins moving along the tangent
$y=ax+const$. The negative sign of the last term in Eq.(\ref{4}) can
be regarded as a manifestation of Le Chatelier principle, i.e. since
an increase in the slope results in the formation of an avalanche,
the velocity components $\dot{x}$ and $\dot{y}$ tend to impede  the
growth of the slope. The positive feedback between $\dot{x}$ and
$y^{\prime}$ in Eq.(\ref{5})
plays an important part in the problem. As we shall see later, it is
precisely the reason behind the self--organization that brings about
the avalanche generation.

After suitable rescaling, Eqs.(\ref{2}), (\ref{4}), (\ref{5}) can be
rewritten in the form of the well--known Lorenz system:

\begin{mathletters}
\label{6}
\begin{eqnarray}
\dot{u}&=&-u+v,\label{6a}\\
\epsilon\dot{v}&=&-v+uS,\label{6b}\\
\delta~\dot{S}&=&(S_{0}-S)-uv, \label{6c}
\end{eqnarray}
\end{mathletters}
where
$ u\equiv (\tau_{y}/ \tau_{x})^{1/2}(\tau_{S} /D)^{1/2}\dot{x},\,
v\equiv (\tau_{y}/ \tau_{x})^{1/2}(\tau_{S}
/D)^{1/2}\dot{y}/a,\,\text{and}\, S\equiv\break (\tau_{y}/
\tau_{x})^{1/2}y^{\prime}/a $ are the dimensionless velocity
components and the slope, respectively;
$\epsilon\equiv \tau_{y}/
\tau_{x},\, \delta\equiv \tau_{S}/ \tau_{x}$  and the  dot now stands for
the derivatives with respect to the dimensionless time $t/\tau_{x}$.
In general, the system
(\ref{6a}) -- (\ref{6c})
cannot be solved analytically, but
in the simplest case, where $\epsilon\ll 1\, \text{and}\, \delta\ll
1$, the vertical velocity $v$ and the slope $S$ can be eliminated by
making use of the adiabatic approximation that implies neglecting of
the left hand sides of Eqs.(\ref{6b}), (\ref{6c}). As a result, the
dependencies of $S$ and $v$ on the horizontal velocity $u$ are given
by

\begin{equation}
S=\frac{S_{0}}{1+u^{2}},\qquad v=\frac{S_{0}u}{1+u^{2}}.
\label{7}
\end{equation}
Note that, under $u$ is in the physically meaningful range between $0$
and $1$, the slope is a monotonically decreasing function of $u$,
whereas the velocity $v$ increases with $u$
(at $u>1$ we have $\text{d}v/\text{d}u<0$ and the flow of the sand
becomes turbulent).

Substituting second Eq.(\ref{7}) into Eq.(\ref{6a}) yields the
Landau--Khalatnikov equation:

\begin{equation}
\dot{u}=-\frac{\partial E}{\partial u}
\label{8}
\end{equation}
with the kinetic energy given by
\begin{equation}
E=\frac{1}{2}u^{2}-
\frac{1}{2}S_{0}\ln{\left(1+u^{2}\right)}.
\label{9}
\end{equation}
For $S_{0}<1$, the $u$--dependence of $E$ is monotonically
increasing and the only stationary value of $u$ equals zero,
$u_{0}=0$, so that there is no avalanches in this case. If the slope
$S_{0}$
exceeds the critical value, $S_{c}=1$, the kinetic energy assumes the
minimum with non--zero steady state
velocity components $u_{e}=v_{e}=(S_{0}-1)^{1/2}$ and
the slope $S_{e}=1$.

The above scenario represents supercritical regime of the avalanche
formation and corresponds to the second--order phase transition. The
latter can be easily seen from the expansion of the kinetic energy
(\ref{9}) in power series of $u^{2}\ll 1$. So the critical
exponents are identical to those obtained within the framework of
the mean field theory [11].

The drawback of the outlined approach is that it fails to account for
the subcritical regime of the self--organization that is the reason
for the appearance of avalanches  and analogous to the
first--order phase transition, rather than the second--order one.
So one has to modify the above theory by taking the assumption that
the effective relaxation time $\tau_{x}(x)$
increases with velocity $u$ from value $\tau_{x}(1+m)^{-1}$,
$m>0$ to  $\tau_{x}$ [12].
The simplest two--parameter approximation  is

\begin{equation}
\frac{\tau_{x}}{\tau_{x}(u)}=1+\frac{m}{1+(u/u_{0})^{2}}
\label{10}
\end{equation}
where $0<u_{0}<1$. The expression for the
kinetic energy (\ref{9}) then changes by adding the term

\begin{equation}
\Delta E=\frac{m}{2}u_{0}^{2}
\ln{\left(1+\left(\frac{u}{u_{0}}\right)^{2}\right)}
\label{11}
\end{equation}
and the stationary values of $u$ are

\begin{eqnarray}
u_{e}^{m}&=&u_{00}
\left(
1\mp \left[
1+u_{0}^{2}u_{00}^{-4}(S_{0}-S_{c})
\right]^{1/2}
\right)^{1/2}, \label{12}\\
2u_{00}^{2}&\equiv& (S_{0}-1)+S_{c}u_{0}^{2},\qquad
S_{c}\equiv1+m.\nonumber
\end{eqnarray}
The upper sign in the right hand side of Eq.(\ref{12}) is for the
value at the unstable state $u^{m}$ where the kinetic energy
$E+\Delta E$ has the maximum, the lower one corresponds to the stable
state $u_{e}$. The corresponding value of the stationary slope

\begin{equation}
S^{m}=\frac
{1+u_{00}^{2}+\sqrt
{\left(1+u_{00}^{2}\right)^{2}-
\left(1-u_{0}^{2}\right)S_{0}}}
{1-u_{0}^{2}}
\label{13}
\end{equation}
smoothly increases from the value

\begin{equation}
S_{\text{min}}=1+u_{0}\sqrt{m/(1-u_{0}^{2})}
\label{14}
\end{equation}
at the parameter $S_{0}=S_{c0}$ with

\begin{equation}
S_{c0}=\left(1-u_{0}^{2}\right)S_{\text{min}}^{2}
\label{15}
\end{equation}
to the marginal value $S_{c}=1+m$ at $S_{0}=S_{c}$. The
$S_{0}$--dependencies of $u_{e},\,u_{m},\,\text{and}\,S_{e}$
are presented in
Fig.1. As is shown in Fig.1a, under the adiabatic condition
$\tau_{S}\ll\tau_{x}$ is met and the parameter $S_{0}$ slowly
increases being below $S_{c}$ ($S_{0}\le S_{c}$), no avalanches can
form. At the point $S_{0}=S_{c}$ the velocity $u_{e}$ jumps upward
to the value $\sqrt{2}u_{00}$ and its further smooth
increase is determined
by Eq.(\ref{12}). If the parameter $S_{0}$ then goes downward the
velocity $u_{e}$ continuously decreases up to the point, where
$S_{0}=S_{c0}\,\text{and}\,u_{e}=u_{00}$. At this point the velocity
jump--like goes down to zero. Referring to Fig.1b, the stationary
slope $S_e$ shows a linear increase from $0$ to $S_{c}$ with the parameter
$S_{0}$ being in the same interval and, after the jump down to
the value $(1-u_{0}^{2})^{-1}$ at $S_{0}=S_{c}$, $S_{e}$ smoothly
decays to $1$ at
$S_{0}\gg S_{c}$. Under the parameter $S_{0}$ then decreases from
above $S_{c}$ down to $S_{c0}$ the slope grows.
When the point (\ref{15}) is reached, the avalanche
stops, so that the slope undergoes the jump from the value
(\ref{14}) up to the one defined by Eq.(\ref{15}). For
$S_{0}<S_{c0}$ again the parameter
$S_{e}$ does not differ from $S_{0}$.
Note that this subcritical regime is realized provided the parameter
$m$, that enters the dispersion law (\ref{10}), is greater than
\begin{equation}
m_{\text{min}} = \frac{u_{0}^{2}}{1-u_{0}^{2}}.
\label{16}
\end{equation}

Clearly, according to the picture described, the avalanche generation
is characterized by the well pronounced hysteresis, when the grains
of sand initially being at rest begin to move downhill only if the
slope of the surface exceeded its limiting value $S_{c}=1+m$, whereas
the slope $S_{c0}$ needed to stop the avalanche is less
than $S_{c}$ (see Eqs.(\ref{14}), (\ref{15})). This is the case in the
limit $\tau_{S}/\tau_{x}\to 0$ and the hysteresis loop shrinks with
the growth of the adiabaticity parameter
$\delta\equiv\tau_{S}/\tau_{x}$.  In addition to the smallness of
$\delta$, the adiabatic approximation implies the ratio
$\tau_{y}/\tau_{x}\equiv\epsilon$ is also small. In contrast to the
former, the latter does not seem to be realistic for the system under
consideration, where, in general, $\tau_{y}\approx\tau_{x}$. So it is
of interest to study to what extent the finite value of $\epsilon$
could change the results.

Owing to the condition $\delta\ll 1$, Eq.(\ref{6c}) is still algebraic and
$S$ can be expressed in terms of $u$ and $v$. As a result, we derive
the system of two nonlinear differential equations that can be
studied by the phase portrait method [12]. The phase portraits
for various values of $\epsilon$ are displayed in Fig.2, where the
point O represents the stationary state and the point S is related to
the maximum of the kinetic energy. As is seen from the figure,
independently of $\epsilon$, there is the universal section that
attracts all phase trajectories and its structure is appeared to be
almost insensitive to changes in $\epsilon$. Analysis of time
dependencies $v(t)$ and $u(t)$ reveals that the velocity components
slow down appreciably on this section in comparison to the rest parts
of trajectories that are almost rectilinear (it is not difficult to
see that this effect is caused by the smallness of $\delta$). Since
the most of time the system is in vicinity of this universal section,
we arrive at the conclusion that finite values of $\epsilon$ do not
affect qualitatively the above results obtained in the adiabatic
approximation.

\section{Noise influence}

To take into account additive noises of the velocity components
$u$, $v$ and the slope $S$ it needs to add
to right-hand parts of Eqs.(\ref{6a}) -- (\ref{6c}) the stochastic terms
$I_u^{1/2}\xi$, $I_v^{1/2}\xi$, $I_S^{1/2}\xi$, respectively
(here the noise intensities $I_{u,v,S}$ are measured in units
$(\tau_{x}/ \tau_{y})(D/\tau_{S})$,
$a^2(\tau_{x}/ \tau_{y})(D/\tau_{S})$,
$a^2(\tau_{x}/ \tau_{y})$, correspondingly, and
$\xi(t)$ is $\delta$-correlated stochastic function) [13].
Then, within the adiabatic approximation
Eq.(\ref{8}) acquires the stochastic addition
$$
\left\{I_u^{1/2} + \left[ I_v^{1/2} g_v(u) + I_S^{1/2} g_S(u)\right]
\right\}\xi(t),
$$
where we introduce the multiplicative functions
$g_S(u)=u g_v(u) = u/(1+u^2)$.
As a result the extreme points of the distribution
$P(u)\propto \exp\lbrace-U(u)\rbrace$
of the stochastic variable $u$ is given by the effective
energy [14]
\begin{equation}
U(u)=\ln I(u)+\int{\partial
E/\partial u\over I(u)}{\rm d}u \label{VI.5_17}
\end{equation}
where the bare energy $E$ is determined by Eq.(\ref{9}) and
the expression for the effective noise intensity
\begin{equation}
I(u)\equiv I_u + I_v g^2_v(u) + I_S g_S^2(u)\label{VI.5_18}
\end{equation}
follows from the known property of additivity
of squares of variances of independent Gaussian random
quantities [13].
Combining expressions
(\ref{9}), (\ref{VI.5_17}), (\ref{VI.5_18}),
we can find the explicit form of $U(u)$,
which is too cumbersome to be reproduced here.
Much simpler is the equation
\begin{equation}
x^3 - S_0 x^2 - 2I_S x + 4(I_S - I_v) = 0,
\quad x \equiv 1 + u^2, \label{VI.5_19}
\end{equation}
which defines the locations of the maxima of distribution function $P(u)$.
According to Eq.(\ref{VI.5_19}), they are insensitive to changes
in the intensity of noise $I_u$ of the velocity component $u$,
but are determined by the stationary value $S_0$
of the sandpile slope and the intensities
$I_v$, $I_S$ of the noises of vertical velocity component $v$
and slope $S$, which acquire the multiplicative character
in Eq.(\ref{VI.5_18}).
Hence, it can put for simplicity $I_u=0$
and Eqs.(\ref{9}), (\ref{VI.5_17}), (\ref{VI.5_18}) give
the follow expression for the effective energy:
\begin{eqnarray}
U(u)&=&{1\over 2}\left[{u^4\over 2} +
(2-S_0-i)u^2+\right. \label{VI.5_27}\\
&&\left. (1-i)\left(1-S_0-i
\right)\ln(i+u^2)\right]+
I_S\ln[g_S^2(u)+i g_v^2(u)], \quad
i\equiv{I_v/I_S}.
\nonumber
\end{eqnarray}

According to Eq.(\ref{VI.5_19}),
the effective energy (\ref{VI.5_27}) has a minimum at $u=0$
if the stationary slope $S_0$
does not exceed the critical level
\begin{equation}
S_c=1+2 I_S-4I_v,\label{VI.5_29}
\end{equation}
whose value increases at
increasing intensity of noise of the sandpile slope, but
decreases with one of the velocity.
In this case, sand grains are at rest.
In the simple case $I_v=0$, the avalanche creation
corresponds to solutions
\begin{equation}
u^2_\pm={1\over
2}\left[S_0-3+\sqrt{(3-S_0)^2+4(2S_0-3+2 I_S)}\right],
\label{VI.5_34}
\end{equation}
which are obtained from Eq.(\ref{VI.5_19}) after
eliminating the root $u^2=0$.
The magnitude of this solution has its minimum
\begin{equation}
u_c^2={1\over 2}\left[
(S_0-3)-\sqrt{(S_0 +7)(S_0-1 )}\right]
\label{VI.5_35}
\end{equation} on the line defined by expression
(\ref{VI.5_29}) with $I_v=0$. At $S_0 < 4/3$,
the roots $\pm u_c$ are complex,
at $S_0 =4/3$ they become zero, at $S_0 >4/3$ they are real,
and  $ u_+=- u_-$.
In this way, the tricritical point
\begin{equation} S_0
= 4/3,\qquad  I_S = 1/6 \label{VI.5_36}
\end{equation}
corresponds to the appearance of roots $ u_\pm\ne 0$ of Eq.(\ref{VI.5_19})
(avalanche creation).
If condition (\ref{VI.5_29}) is satisfied,
the root $ u=0$ corresponds to the minimum of the
effective energy (\ref{VI.5_27})
at $S_0 < 4/3$, whereas at $S_0 > 4/3$
this root corresponds to the maximum, and the roots   $ u_\pm$ to
symmetrical minima.

           Now we find another condition of stability of roots
$ u_\pm$.  Setting the discriminant of Eq.(\ref{VI.5_19}) equal to
  zero, we get the equations
\begin{equation}
I_S=0,\qquad I_S^2-I_S
\left[{27\over 2}\left(1-{S_0\over 3}\right)-{S_0^2\over 8}\right]+{S_0^3\over 2}=0,
\label{VI.5_37}
\end{equation}
the second of which gives
\begin{equation}
2I_S=\left[{27\over 2}\left(1-{S_0\over 3}\right) -{S_0^2\over
8}\right] \pm  \left\{ \left[ {27\over 2}\left(1-{S_0\over
3}\right) -{S_0^2\over 8}\right]^2-
2S_0^3\right\}^{1/2}.\label{VI.5_38}
\end{equation}
This equation
defines a bell-shaped curve $S_0(I_S)$, which intersects with
the horizontal axis at the points $I_S=0$  and
$I_S=27/2$, and has a maximum at
\begin{equation}
S_0=2, \qquad I_S=2.  \label{VI.5_39}
\end{equation}
It is easy to see that for $I_v=0$
this line touches the curve (\ref{VI.5_29}) at point
(\ref{VI.5_36}).

        Let us now consider the more general case of two
multiplicative noises $I_v, I_S\ne 0$.
Introducing the parameter $a=1-i$, $i=I_v/I_S$ and the renormalized
variables $\widetilde{I}\equiv {I_S/ a^2}$,
$\widetilde{S_0}\equiv{S_0/ a}$, $ \widetilde{u}^2 =
(1+ u^2)/a-1$, at $i<1$ we
may reproduce all above expressions
with the generalized energy $\widetilde{U}/\widetilde{I}$
in Eq.(\ref{VI.5_27}).
Then the action of the noise of the vertical component
of the velocity $v$ is reduced to the
renormalization of the extremum value of the horizontal one
by the quantity $(a^{-1}-1)^{1/2}$, so that the region of divergence
$\widetilde{u}\approx 0$ becomes
inaccessible.

      The condition of extremum of the generalized energy
(\ref{VI.5_27}) splits
into two equations, one of which is simply  $ u =0$, and the other
is given by Eq.(\ref{VI.5_19}). As pointed out above,
analysis of the latter indicates
that the line of existence of the zero solution is defined by an
expression (\ref{VI.5_29}).
The tricritical point  has the coordinates
\begin{equation} S_0 ={4\over 3}(
1- I_v),\quad  I_S ={1\over 6}\left( 1
+8 I_v\right). \label{VI.5_48}
\end{equation}
The phase diagram for the fixed intensities $I_v$ is shown in Fig.3.
Here the curves $1$, $2$ define the thresholds
of absolute loss of stability
for the fluxless and flux steady-states, respectively.
Above line $1$ the system occurs in a stable flux state,
below curve $2$ it is in fluxless one, and between these lines
the two-phase domain is realized.
For $ I_v<1/4$ situation is generally the same as
in the simple case $I_v=0$ (see Fig.3a).
At  $I_v>1/4$ the SOC regime is possible for small intensities $I_S$
of the slope noise (Fig.3b).
According to (\ref{VI.5_48}) the tricritical point
occurs on the  $I_S$ axis at
$ I_v=1$, and for the noise intensity $I_v$
larger than the critical value $I_v=2$
the stable fluxless state disappears (see Fig.3c).

\section{Conclusion}

According to the above consideration, the dissipative dynamic
of grains flow in a real sand pile can be represented
within the framework of Lorenz model,
where the horizontal and vertical velocity components
play a role of an order parameter and
its conjugate field, respectively,
and the sandpile slope is a controlling parameter.
In Sec.2, the noiseless case is examined to show
that an avalanche creates
if the externally driven sandpile slope
$y'_0$ is larger than the critical magnitude
$\gamma(\tau_x \tau_y)^{-1/2}$.
In this sense, the systems
with small values of the kinetic coefficient $\gamma$
and large relaxation times $\tau_x$, $\tau_y$
of the velocity components are preferred.
However, the sand flow appears here as a phase transition
because the spontaneous avalanche creation
is impossible in the noiseless case.
Taking into account the additive noises
of the above degrees of freedom,
we show in Sec.3 that the stochasticity influence
is non-essential for the horizontal velocity component
and is crucial for the vertical one.
The SOC appears if the noise intensity
of the latter exceeds the value
$2^{-2}(D\gamma^2/\tau_x\tau_y\tau_S)$, provided the noise is small
for the sandpile slope.
If the noise intensity
of the vertical velocity component is larger than the critical value
$2(D\gamma^2/\tau_x\tau_y\tau_S)$,
the fluxless steady-state disappears at all.


\newpage
\begin{center}
{\bf FIGURE CAPTIONS}
\end{center}

Fig.1. The $S_{0}$--dependencies of a) the velocities $u_{e}$,
$u_{m}$,
and b) the equilibrium slope $S_{e}$. The arrows indicate
the hysteresis loop.

Fig.2.
Phase portraits in the $v-u$ plane at $S_{0}=1.25S_{c0}$
for a) $\epsilon=10^{-2}$; b) $\epsilon=1$; c) $\epsilon=10^{2}$.

Fig.3.
Phase diagrams for fixed values $I_v$ of the noise intensities
of the vertical velocity component:
a) $I_v=0$, b) $I_v=1$, c) $I_v=2$. Curves 1 and 2 define the
boundary of stability of avalanche and non-avalanche phases; A --
avalanche phase, N -- non-avalanche.

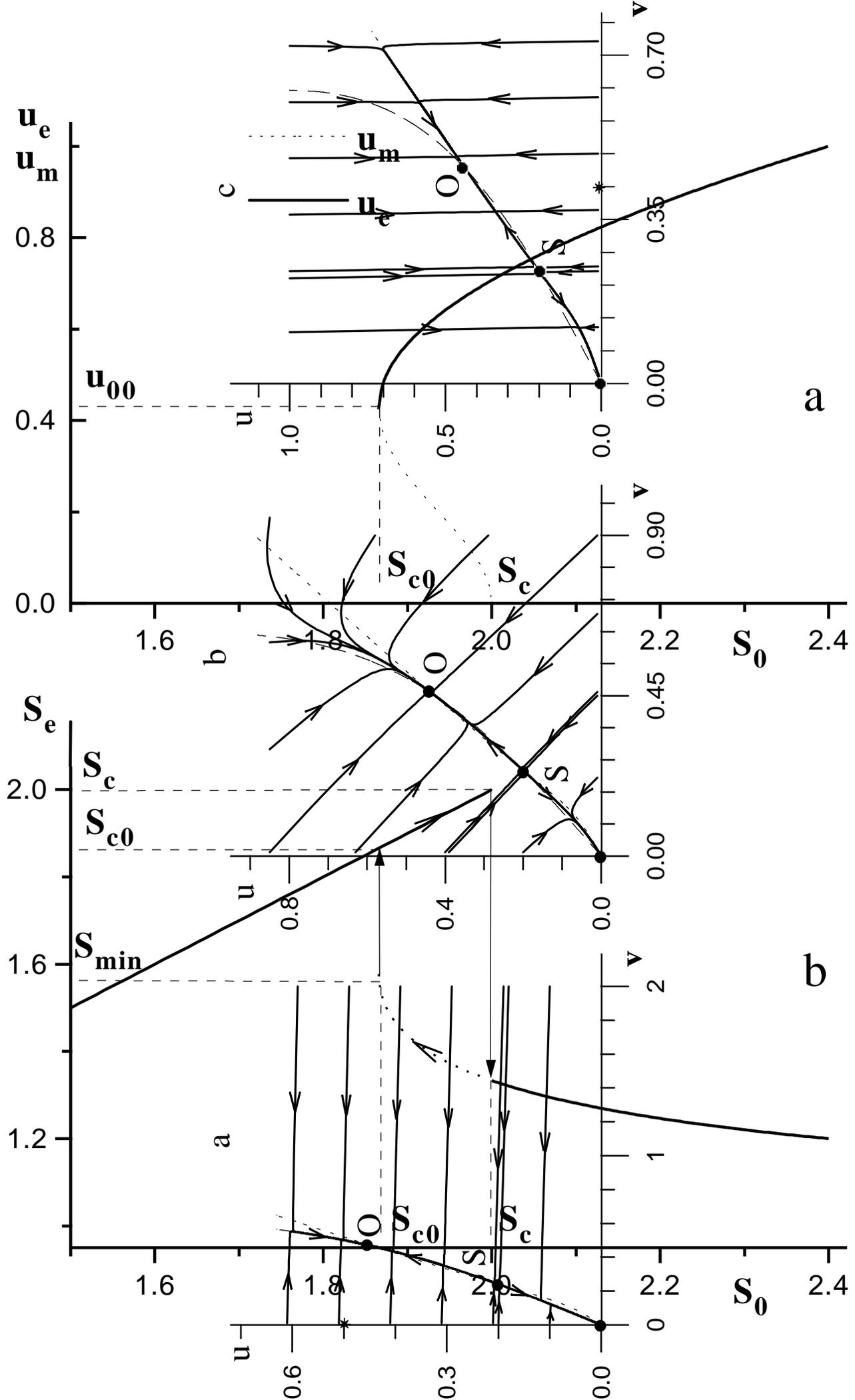
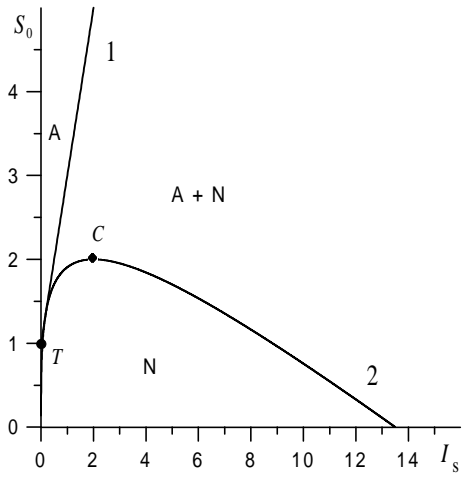
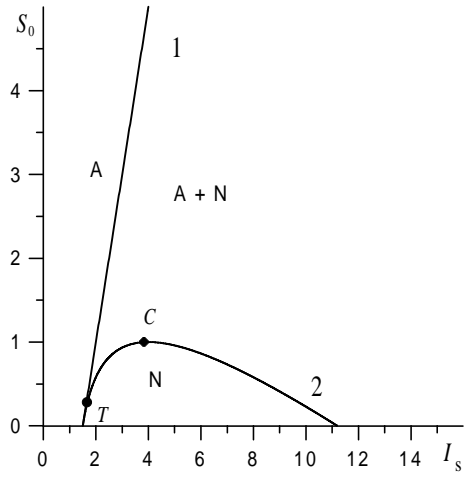
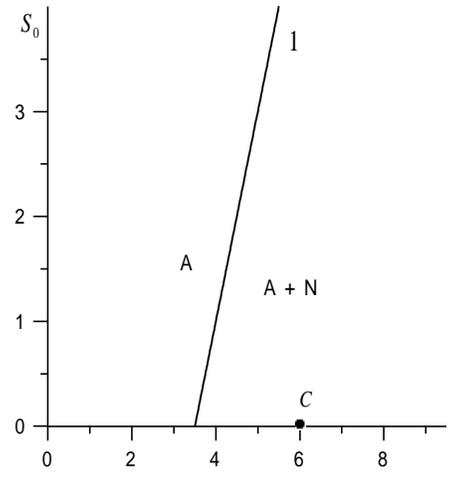
\end{document}